\documentstyle[emulateapj,psfig]{article}

\input psfig.sty

\def\simgt{\lower.5ex\hbox{$\; \buildrel > \over \sim \;$}}
\def\simlt{\lower.5ex\hbox{$\; \buildrel < \over \sim \;$}}
\def\sp{\hspace{1.5pt}}

\def\src{{1E\sp1841$-$045}}
  
\def\sp{\hspace{1.5pt}}
\def\etal{{et~al.}}
\def\amin{\ifmmode^{\prime}\else$^{\prime}$\fi}
\def\asec{\ifmmode^{\prime\prime}\else$^{\prime\prime}$\fi}

\def\simgt{\lower.5ex\hbox{$\; \buildrel > \over \sim \;$}}
\def\simlt{\lower.5ex\hbox{$\; \buildrel < \over \sim \;$}}

\newcommand\xte{{\it RXTE}}

\newcommand\ginga{{\it Ginga}}

\newcommand\sax{{\it Beppo}SAX}
\newcommand\SAX{{\it Beppo}SAX}
\newcommand\ASCA{{\it ASCA}}
\newcommand\asca{{\it ASCA}}

\newcommand\rosat{{\it ROSAT}}
\def\sp{\hskip 1.5pt}

\def\kes73{\hbox{Kes\sp73}}

\lefthead{Gotthelf, Vasisht \& Dotani}
\righthead{Spin-down of the \kes73\ Pulsar}

\begin{document}

\title{On the Spin History of the X-ray Pulsar in Kes\sp73: \\
Further Evidence For an Utramagnetized Neutron Star}

\author{E. V. Gotthelf$^1$, G. Vasisht$^2$, \&\ T. Dotani$^3$}
\altaffiltext{1}{Columbia Astrophysics Laboratory, Columbia University, 550 West 120$^{th}$ Street, New York, NY 10027, USA; evg@astro.columbia.edu} 
\altaffiltext{2}{Jet Propulsion Laboratory, California  Institute of Technology, 4800 Oak Grove Drive, Pasadena, CA, 91109, USA; gv@astro.caltech.edu}
\altaffiltext{3}{Institute of Space and Astronautical Science, 3-1-1 Yoshinodai, Sagamahara, Kanagawa 229, Japan; dotani@astro.isas.ac.jp}

\begin{abstract}

In previous papers, we presented the discovery of a 12-s X-ray pulsar
in the supernova remnant \kes73\, providing the first direct evidence
for an ultramagnetized neutron star, a magnetar, with an equivalent
dipole field of nearly twenty times the quantum critical magnetic
field ($m_e^2 c^3 / e \hbar$). Our conclusions were based on two epochs of measurement of the
spin, along with an age estimate of the host supernova remnant.
Herein, we present a spin chronology of the pulsar using additional
\ginga, \asca, \xte, \& \sax\ datasets spanning over a decade. Timing
and spectral analysis confirms our initial results and severely limit
an accretion origin for the observed flux. Over the 10 year baseline,
the pulsar is found to undergo a rapid, constant spindown, while
maintaining a steady flux and an invariant pulse profile. Within the
measurement uncertainties, no systematic departures from a linear
spin-down are found - departures as in the case of glitches or simply
stochastic fluctuations in the pulse times-of-arrival (e.g.  red
timing noise). We suggest that this pulsar is akin to the soft
$\gamma$-ray repeaters, however, it is remarkably stable and has yet
to display similar outbursts; future $\gamma$-ray activity from this
object is likely.

\end{abstract}

\keywords{stars: individual (\kes73, \src) --- stars: neutron --- 
supernova remnants --- X-rays: stars}

\section{Introduction}

The discovery of 12-s pulsed X-ray emission from the compact source
within \kes73\ (Vasisht \& Gotthelf 1997; herein VG97, Gotthelf \&
Vasisht 1997) came somewhat as a surprise, as this Einstein source
(\src) had been studied for some time (Kriss \etal\ 1985; Helfand et
al. 1994).  The pulsar was initially detected in an archived \asca\
observation (1993) of \kes73\ (also SNR G27.4+0), and soon confirmed in
an archived \rosat\ dataset.  The measured spindown from these
detections indicated rapid braking on a timescale of $\tau_s \simeq
4\times 10^3$ yr, consistent with the inferred age of the supernova
remnant. The similarity in age along with the geometric location of
the pulsar in the center of the symmetric and well defined remnant
strongly suggests that the two objects are related.

There is sufficient evidence to argue that the host supernova remnant
\kes73\ is relatively young, at age $\sim 2\times 10^3$ yr (Helfand et
al. 1994; Gotthelf \& Vasisht 1997).  Morphologically, it resembles
any classic, limb-brightened shell-type radio supernova remnant, $\sim
4'$ in diameter, located at an HI derived distance of $6.0-7.5$ kpc
(Sanbonmatsu \& Helfand 1992). Its diffuse X-ray emission is
distributed throughout the remnant and has a spectrum characteristic of
a hot plasma, $kT \simeq 0.8$ keV, along with fluorescence lines of
several atomic species, including O-group elements, that indicate a
young blast-wave of Type II or Ib origin, still rich in stellar
ejecta.  The relative abundance of ionized species of Si and S
observed in the \kes73\ spectrum, suggest a level of ionization
in-equilibrium consistent with an age $\simlt 2\times 10^3$ yr.

Based on the observed characteristics of \kes73\ and its X-ray pulsar,
we suggested (VG97) that \src\ was similar to the `anomalous' X-ray
pulsars (Mereghetti \& Stella 1995; van Paradijs, Taam \& van den
Heuvel 1995). We argued that \src\ could not be an accreting neutron
star; this was based on evolutionary arguments and the relative youth
of \kes73\ and the fact that we found no evidence for accretion in our
datasets. Instead, it was proposed that the X-ray pulsar was powered
by an ultramagnetized neutron star with a dipole field of $B_s \simeq
7\times 10^{14}$ G, and was the first of its kind; magnetic braking
was assumed to be the predominant spindown mechanism, with $B_s\propto
(P\dot P)^{1/2}$.  Since then, large magnetic fields have also been
inferred for the soft $\gamma$-ray repeaters 1806$-$20 and 1900$+$14
(see Kouveliotou \etal\ 1998, 1999) via measurement of their spins and
spindown with \xte\ and \asca.

In this paper we present a longterm spin history of the X-ray
pulsar. We reinforce our original \ASCA\ and \rosat\ datasets with new
\ASCA, \xte, \& \SAX\ pointings and ten year-old \ginga\ archival
datasets, bringing the total number of timing observations of \kes73\ to
seven. We show that the spin evolution obeys a steady linear spindown
at a rate consistent, within errors, with our original estimate;
re-affirming our somewhat marginal \rosat\ detection (VG97).

\section{Observations}

\subsection{\ASCA\ \& \SAX\ Datasets}

Kes\sp73 was re-observed with \ASCA\ (Tanaka \etal\ 1994) on March
27-28, 1998, using an observing plan identical with the original 1993
observations. A complete description of the observing modes can be
found in VG97.  Here we concentrate on the high temporal resolution
data ($62 \ \mu$s or 48.8 ms depending on data mode) acquired with the
two Gas Imaging Spectrometers (GIS2 \& GIS3) on-board \asca. The
datasets was edited using the standard Rev 2 screening criteria
which resulted in an effective exposure time of 39 ks per GIS
sensor. Photons from the two GISs were merged and arrival times
corrected to the barycenter. A log of all observations presented
herein are given in Table 1.

We also acquired a 1.5 day \sax\ observation of \kes73\ on March 8-9,
1999 using the three operational narrow field instruments, the Low Energy
Concentrator (LECS) and two Medium Energy Concentrators (MECS2 \&
MECS3).  These imaging gas scintillation counters are similar to the
GIS detectors, providing arcminute imaging over a $\sim
40^\prime$ field-of-view in a broad energy band-pass of $0.1 - 12$ kev
(LECS) and $1 - 12$ kev (MECS), with similar energy resolution; the
data consists of photon arrival times tagged with 16 $\mu$s
precision. All data were pre-screened during the standard SAX pipeline
processing to remove times of enhanced background resulting in usable 
exposure times of 57.9 ks for each of the MECSs and 26.5 ks for
the LECS. Here we concentrate exclusively on photons obtained with the
two MECS detectors, which makes up the bulk of the data.
Data from the two MECS were merged and barycentered by the SAX team.

For all data sets we extracted barycenter corrected arrival times from
a $\simeq 4'$ radius aperture centered on the central object in
\kes73\ and restricted the energy range to $2-10$ keV. We then
searched these time series for coherent pulsation by folding the data
about the expected periods derived from the ephemeris of VG97.  In
each case highly significant power is detected in the resulting
periodograms corresponding to the central pulsar's pulse period at the
specific epoch. Figure 1 compares periodograms of the \asca\ data of
1993 and 1998 produced in the manner described above. We plot these on
the same scale to emphasized both the significance of the detection
and the unambiguous change in period between the two epochs.

\subsection{\ginga\ \& \xte\ Datasets}

With a period detection in hand, we re-analysed archival data from the
\ginga\ (Makino 1987) and \xte\ (Bradt \etal\ 1993) missions.  The
main instrument on-board \ginga\ is the non-imaging Large Area Counter
(LAC) which covers an energy range of $1-37$ kev with an effective
area of $4000$ cm$^2$ over its $\sim 2 \times 2$ field-of-view.
\ginga\ observed \kes73\ twice in data modes with sufficient temporal
resolution ($2 \ \mu$s or $16 \ \mu$s depending on data mode) and
exposure to carry out the present analysis (see Table 1). These data
were screened using the following criteria: i) Earth-limb elevation
angle $ >5$ degrees ii) cut-off rigidity $> 8$ GeV/c, and iii) South
Atlantic Anomaly avoidance. The LAC light curves were restricted to
the $\sim 1 - 17$ keV energy band-pass and corrected to the solar
heliocenter using available software\footnote{The barycentric
correction to these periods is small and has been added in, along with
the statistical error in the periods.}

Kes\sp73 was observed for 5 ks by \xte\ during 1996 as part of the GO
program. We analyzed archive data acquired with the Proportional
Counter Array (PCA) in ``Good Xenon'' data mode at 0.9 $\mu$s time
resolution. The PCA instrument is similar to the LAC, with a smaller
field-of-view and a greater effective area of $6,500$ cm$^2$. The two
\xte\ observation windows were scheduled so that no additional time
filtering was required. After processing and barycentering the Good
Xenon data according to standard methods, we selected events from
layer 1 only and applied an energy cut of $\simlt 10$ keV.

As found with the imaging data, epoch folding around the anticipated
period produced a highly significant period detection. These periods
are consistent with the extrapolated \asca\ derived ephemeris. In
Figure 2 we display pulse profiles of \kes73\ at two epochs separated
by over a decade to look for possible long term changes. To
improve the signal-to-noise in the latter observation, we have
co-added phased aligned profiles from the 1998 \asca\ and the 1999
\sax\ observations.  No significant differences were found between the
two pulse profiles, which are identical to the 1993 \asca\ and 1996
\xte\ profiles, to within statistical uncertainties.

\subsection{\src: Timing Characteristics}

In order to accurately determine the detected period at each epoch we
oversampled the pulse signal by zero-padded the lightcurves (binned at
1 s resolution) to generate $2^{20}$ point FFTs. We then fit for the
centroid to the peak signal to determine the best period. In none of
the cases do we detect significant higher harmonics, a fact consistent
with the roughly sinusoidal shape of the pulsar's profile.

To estimate the uncertainty in the period measurements we carried out
extensive Monte-Carlo simulations. For each data set we generated a
set of simulated time series whose periodicity, total count rate, and
noise properties and observation gap are consistent with the actual
data set for each epoch.  We used the normalized profile folded into
10 bins, to compute the probability of a photon arriving in a given
phase bin. Each realization of the simulated data was subjected to the
same analysis as the actual data sets to obtain a period
measurement. After 500 trials we accumulated a range of
measured periods which was well represented by a normal
distribution. The resulting standard deviation of this function is
taken as the 1-sigma uncertainty in the period and is presented in table
1. The errors in the period are roughly consistent with the size of a
period element divided by the signal-to-noise of each detection.

Each period measurement was assigned an epoch defined as the
mid-observation time (in MJD) for that data set. The period measurements
and their uncertainties were then fit with simple first-order and
second-order models to evaluate the spindown characteristics.  The
parameters were $P_s$, $\dot P_s$ with the addition of the second
derivative $\ddot P_s$ for the second order fit, with the spin history
written as a Taylor expansion.  The best fit to the linear model gives
the following period ephemeris (Epoch MJD 49000), $P = 11.765732 \pm\
0.000024$ s; $\dot P = 4.133 \times 10^{-11} \pm\ 1.4 \times 10^{-13}$
s s$^{-1}$.

The linear spindown model was consistent with the data with fit
residuals at the $10^{-4}$ s (or 0.1 micro-Hz) level. We found that
these residual were not sensitive to a second derivative of the
period. An upper-limit allowed by the available datasets and
their associated errors is more than an
order-of-magnitude larger than expected just due to classic vacuum
dipole spindown; for a vacuum dipole rotator one expects $\ddot P_s
\simeq (2 - n) {\dot P_s^2} / P_s$; where $n = 3$ is the braking index
in the vacuum dipole formalism.

\section{Discussion}

In its observed characteristics, \src\ most resembles the anomalous
X-ray pulsars (AXPs), with its slow, $\sim 10$ s pulse period, steep
X-ray spectral signature, inferred luminosity of $\sim 4 \times
10^{35}$ erg s$^{-1}$ cm$^{-2}$, and lack of counterpart at any
wavelength. It is, however, unique among these objects in its apparent
temporal and spectral stability. The two AXPs for which sufficient
monitoring data is available, 2259+586 and 1E 1048$-$59, show large
excursions in flux ($\simgt 3$) and significant irregularities in
their spin down ($ \log(| \delta P/P |) \sim -4$).  Compared with
these objects, the spindown of \src\ suggests a lower level of torque
fluctuations.  For \src, the magnitude of timing irregularities, given
by the timing residuals after subtracting the linear model, is $\log
(|{\sigma /P}|) < -5$, where $\sigma \sim 10^{-4}$ s is the typical
size of the measurement error (Table 1).  The spindown of this object
is apparently quieter than that observed in some middle-aged pulsars,
which have red noise fluctuations in the pulse times-of-arrival of
order $-$3 to $-$4 (e.g. Arzoumanian \etal\ 1994).

There is mounting evidence that, as a class, AXPs are related to the
soft $\gamma$-ray repeaters (SGR), given their similar X-ray spectral
and timing properties.  If the spindown were to show systematic
departures from linearity as in the case of glitches, which might be
accompanied by bursting activity (as is the case in the SGRs, and may
be expected in \src\ if it is ultramagnetized), then such activity has
so-far not been detected by orbiting $\gamma$-ray observatories, nor
is it reflected in the spin history.  Note that the glitch observed
from SGR 1900+14 resulted in a period change an order of
magnitude larger, $ \log(| \delta P/P |) \simeq -4.3$ (Kouveliotou
\etal\ 1999).

In the context of an evolutionary link between the SGRs and AXPs, the
timing and spectral stability of \src\ suggest a quiescent state
either pre- or post- SGR activity. The relative age of \src\ argues
for an early state, possibly preceding $\gamma$-ray
activity, as there is some evidence that \src\ is the youngest amongst
the currently recognized AXPs and SGRs. Half the AXPs are known to be
associated with supernova remnants, while of the four known soft
repeaters - two have host remnants while another, SGR 1806$-$20, has
associated plerion-like emission but no discernible supernova shell.
All these objects are thought be at least $\simgt 10^4$ yr-old, with
the oldest AXPs having been around for a few $\times 10^5$ years
(spindown on timescales of a few hundred years observed in SGRs is no
reflection of their true ages).

Rotational energy loss is insufficient to power the inferred
luminosity of \src, unlike in the usual radio pulsar.  In the SGRs,
the ultimate mechanism for powering particle acceleration is naturally
the release of magnetic free energy (both steady and
episodic-seismic), rather than rotation. The episodic emission of
$\gamma$-ray bursts it thought to be due to starquakes in the neutron
star crust. This could suggest a future ``turning-on'' of \src\ as a
$\gamma$-ray repeater on several thousand year timescale, presumably
resulting from a slow buildup of stress between then core and surface
of the neutron star due to, a yet, unknown state transition in the
stellar crust.

The blackbody spectrum suggests a radiating surface of size $R_\infty
\simeq 8d^2_7$ km, were $d_7$ is the distance to \kes73\ in units of 7
kpc, which is consistent with neutron stellar dimensions (assuming an
isotropic emitter and ignored surface redshift and photospheric
corrections to the observed spectral energy distribution).  This rough
estimate suggests low surface temperature anisotropies (as opposed to
small hotspots), and is in agreement with the low modulation, broad
pulse originating from near the stellar surface.  In an
ultramagnetized neutron star, outward energy transport from a decaying
magnetic field can keep the surface at elevated temperatures, with
high thermal luminosities ($\sim 10^{35}$ erg s$^{-1}$), not observed
normal neutron stars at age $\sim 10^3$ yr. In contrast, researchers
have argued (Heyl \& Hernquist 1998) that the X-ray luminosities in
such stars may be driven by the cooling of the neutron star through a
strongly magnetized, light-element envelope without the need for
appreciable field decay (see also Heyl \& Kulkarni 1999). Along with
these cooling emissions from the surface, the star may have a
magnetically-driven, charged-particle outflow as is suggested by VLA
observations of the SGR 1806$-$20 (Frail, Vasisht \& Kulkarni 1997)
via the energetics and small scale structure of its plerion.  Evidence
a more episodic particle ejection, rather than a steady wind, is
inferred from the recent radio flare observations of SGR 1900+14,
taken during a period of high activity (Frail, Kulkarni \& Bloom
1999).

For \src, we can only attempt to place limits on a steady
pair-wind luminosity: upper limits to radio emission from a putative
plerionic structure surrounding the pulsar, suggest an averaged pair
luminosity of less than $10^{36}$ erg s$^{-1}$. Similarly, limits on a
hard X-ray tail suggest a present day wind luminosity to be less than
$5\times 10^{35}$ erg s$^{-1}$; the latter condition assumes a tail
with a photon index of 2, quite typical of plerions, and a soft X-ray
radiation conversion efficiency of 10 percent. This bounds are well
within the energy budget available from field decay,
$$L_B \simeq (1/6)\dot BB R^3,$$ which is expected to power a
bolometric luminosity of $\simlt 10^{36}$ erg s$^{-1}$; the fastest
avenue for stellar field decay would be the modes of ambipolar
diffusion for which theoretical arguments suggest a timescale of decay
($B /\dot B$) of $\sim 3 B_{15}^2$ kyr (Goldreich \& Reisenegger
1992). Note that these upper limits are a factor $\sim 10^2$ larger
than the star's dipole luminosity of $4\pi^2I\dot P/P^3 \sim 10^{33}$
erg s$^{-1}$. This suggests that spindown torques on the star could
conceivably be dominated by wind torques, although such a wind would
have to be remarkably steady to not produce timing residuals larger
than those observed (Thompson \& Blaes 1998). Alternatively, if
classical dipole radiations is the primary spindown mechanism, then
the stellar magnetic field is $B \simeq 0.75B_{15}$ G.

To conclude, ongoing X-ray timing monitoring of the spin period is
underway by independent groups, including ours, which will provide
accurate measurement of the braking index of this pulsar. As
previously mentioned, a large braking index may directly reflect on
active field decay or field re-alignment inside the the star. The
observed braking index in a pure dipole rotator with loss of torque
due to field decay may be written as $$ n_{obv} = 3 - 2({P\over \dot
P})({\dot B\over B}).$$ Given that the spindown timescale for this
pulsar is about 9 kyr, a field decay time of about $\simlt 3 B_{15}^2$
kyr could impose a fairly large curvature on the spindown.
Perversely, the situation may be far more complicated with different
factors such as dipole radiation, winds driven by magnetic activity,
and field decay all competing for torque evolution. This, however,
remains to be tested through accurate longterm timing. Measuring the
pulsar spin via a series of closely spaced observations may also
reveal a wealth of information on possible glitching and the
subsequent recovery by the star.

\noindent
{\bf Acknowledgments:} EVG is indebted to Jules Halpern and Daniel
Q. Wang for discussions and insights into timing noise and measurement
error. We thank Angela Malizia for barycentering our \sax\ data.
This research is supported by the NASA LTSA grant NAG5-22250.

\begin{deluxetable}{lccccc}
\tablewidth{400pt}
\tablecaption{Observation Log for the \kes73\ Pulsar
\label{tbl-1}}
\tablehead{
\colhead{Start Date} & \colhead{Mission} & \colhead{Exposure} &  \colhead{Epoch$^a$} & \colhead{Period} & \colhead{Uncert.$^b$}  \\
 \colhead{UT}  & \colhead{} &  \colhead{(ks)} & \colhead{(MJD)} & \colhead{(s)} & \colhead{($\mu$s)} 
}
\startdata
08 May \hfil 1987 06:28 & \ginga\ & 37.6 & 46924.5142361 & 11.758320 &  7   \nl
24 Apr \hfil 1991 17:27 & \ginga\ & 11.6 & 48371.2447166 & 11.76346  &  23   \nl
16 Mar \hfil 1992 04:30 & \rosat\ & 25.1 & 48698.1946593 & 11.7645   & \dots \nl
11 Oct \hfil 1993 14:56 & \asca\  & 39.6 & 49272.0820309 & 11.76676  &  45   \nl
31 Aug \hfil 1996 14:21 & \xte\   &  5.9 & 50326.6953400 & 11.7707   & 375   \nl
27 Mar \hfil 1998 22:30 & \asca\  & 38.7 & 50900.4509668 & 11.77248  &  50   \nl
08 Apr \hfil 1999 03:41 & \sax\   & 57.8 & 51276.8889120 & 11.77387  &  34   \nl
\enddata
\tablenotetext{a}{Epochs are given for the mid-observation, with time in MJD.}
\tablenotetext{b}{Period uncertainties quoted as $1\sigma$; see text for method.}
\end{deluxetable}

\bigskip
\begin{figure}
\epsfxsize=6.0cm
\centerline{\psfig{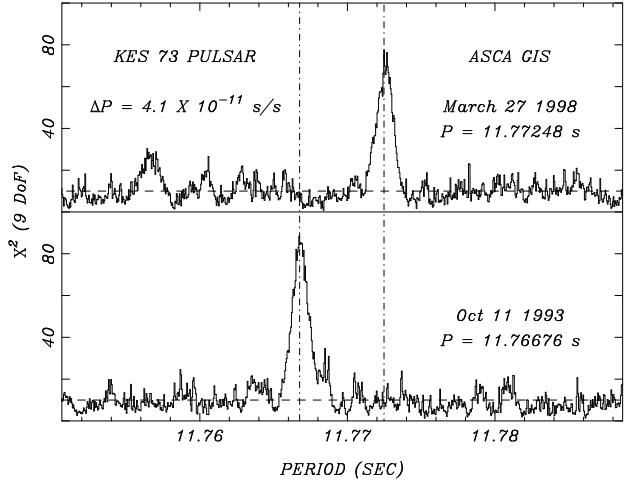}}
\caption{The \asca\ detections of the \kes73\ pulsar at two
epochs. These periodograms depict $\chi^{2}$ of the light curve
folded into 10 phase bins tested against a null-hypothesis as a function of
trial fold periods. (Lower panel) The original discovery detection of
Oct 11 1993 GIS and (Upper panel) the follow-up March 27 1998 GO
observation. The change in period is highly significant and clearly
resolvable; the period derivative, assuming a linear trend, is $\simeq
4.1 \times 10^{-11}$ s/s. The dashed horizontal lines represent the
expected $1\sigma$ noise levels. }
\end{figure}

\bigskip
\begin{figure}
\epsfxsize=6.0cm
\centerline{\psfig{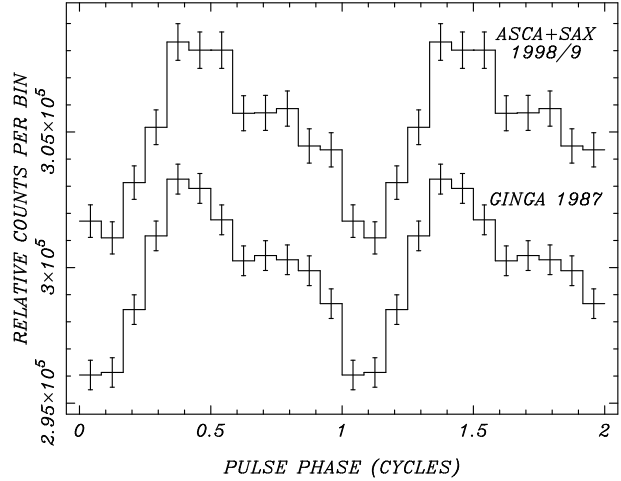}}
\caption{The pulse profiles of the \kes73\ pulsar at two epochs
aligned in phase.  The 1987 \ginga\ data (Bottom), and the
1998 \asca\ and 1999 \sax\ data (Top) phased and co-added to give
increased signal-to-noise. An invariant, Bactrian camel-like profile
is observed in both cases. Two cycles are shown for clarity.}

\end{figure}

\bigskip
\begin{figure}
\centerline{\psfig{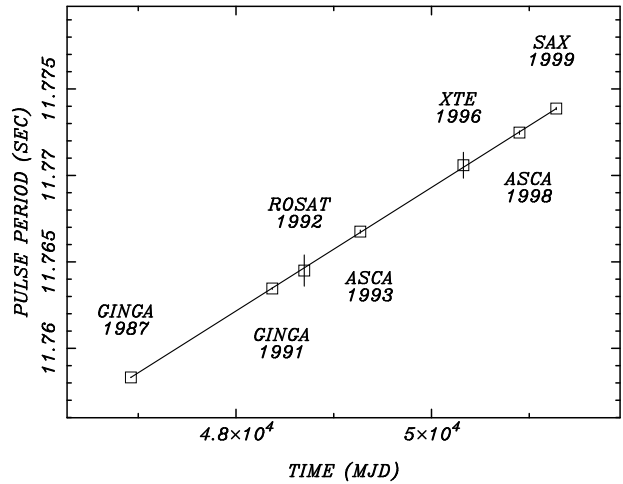}}
\caption{The long-term spin-down history of the \kes73\ pulsar. The square
markers denote the spin period versus epoch; the solid
line follows the best fit linear model (model parameters in section 2)
The data points are depicted with $1\sigma$ error bars.
}
\end{figure}


\begin{thebibliography}{}

\bibitem[Arzoumanian \etal\ 1994]{Arz94} Arzoumanian, Z., Nice, D. J., 
Taylor, J. H., \& Thorsett, S. E. 1994, ApJ, 325

\bibitem[Boella \etal\ 1997]{Boe97} Boella, G., Butler, R. C., Perola,
G. C., Pero, L., Scarsi, L. \& Bleeker, J. A. M. 1997, A\&ASS, 122, 299

\bibitem[Bradt \etal\ 1993]{Brs93} Bradt, H. V., Rothschild, R. E. \& Swank, J. H. 1993, A\&AS, 97, 355.

\bibitem[Corbet \etal\ 1995]{Cor95} Corbet, R. H. D., Smale, A. P.,
Ozaki, M., Koyama, K. \& Iwazawa, K. 1995, ApJ, 443, 786

\bibitem[Cordes \& Helfand 1980]{Ch80} Cordes, J. M. \& Helfand,
D. J. 1980, ApJ, 239, 640

\bibitem[Frail \etal\ 1999]{Frb99} Frail, D. A., Kulkarni,
S. R. \& Bloom, J. S. 1999, Nature, in press; astro-ph/981245

\bibitem[Frail \etal\ 1997]{Fra97} Frail, D. A., Vasisht, G. \& Kulkarni,
S. R. 1997, ApJ, 480, L129                               

\bibitem[Ghosh \etal\ 1997]{Gho97} Ghosh, P., Angelini, L. \& White, N. E.
1997, ApJ, 478, 713

\bibitem[GoldReich 1992]{Gol92} Goldreich, P. \& Reisenegger, A. 1992,
ApJ, 395, 250

\bibitem[Gotthelf \& Vasisht 1997]{Got97} Gotthelf, E. V. \& Vasisht,
G. 1997, ApJ, 486, L129

\bibitem[Helfand \etal\ 1995]{Hel95} Helfand, D. J., Becker, R. H. \&
White, R. L. 1994, ApJ, 434, 627

\bibitem[Heyl \& Hernquist 1998]{HH98} Heyl, J. S. \& Hernquist,
L. 1998, MNRAS, 297, L69

\bibitem[Heyl \& Kulkarni 1999]{HK99} Heyl, J. S. \& Kulkarni,
S. R. 1998, ApJ, 506, L61

\bibitem[Kriss \etal\ 1995]{Kri85} Kriss, G. A., Becker, R. H., Helfand,
D. J. \& Canizares, C. J. 1985, ApJ, 288, 703

\bibitem[Kouveliotou \etal\ 1998]{kou98} Kouveliotou, C. \etal\ 1998,
Nature, 391, 235

\bibitem[Kouveliotou \etal\ 1999]{Kou99} Kouveliotou, C., Strohmayer, T., 
Hurley, K., van Paradijs, J.,
Dieters, S., Woods, P., Thompson, C. \& Duncan, R.C., 1999, astro-ph\/9809140

\bibitem[Makino \etal\ 1987]{Mak87} Makino, F. \etal\ 1987, Astrophys. Letters Commun. 25, 223

\bibitem[Mereghetti \& Stella 1995]{MS95} Mereghetti, S. \& Stella, L. 1995, 
ApJ, 442, L17
\bibitem[Sanbonmatsu \& Helfand 1992]{San92} Sanbonmatsu, K. Y. \&
Helfand, D. J. 1992, AJ, 104, 2189

\bibitem[Parmar \etal\ 1997]{Par97} Parmar, A. N. \etal\ 1997,
A\&A, 323, L29

\bibitem[Tanaka \etal\ 1994]{Tan87} 
Tanaka, Y., Inoue, H. \& Holt, S. S. 1994, PASJ, 46(3), L37

\bibitem[Thompson \& Blaes 1998]{TB98} Thompson, C. \& Blaes, O. 1998, Phys. Rev. D, 57, 3219

\bibitem[van Paradijs, Taam \& van den Heuvel 1995]{vPTvdH95} van Paradijs, J., Taam, R. E. \& van den Heuvel, E. P. J. 1995, A\&A 299, L41
 
\bibitem[Vasisht \& Gotthelf 1997]{vas97} Vasisht, G. \& Gotthelf,
E. V. 1997, ApJ, 486, L129 (VG97)

\end{thebibliography}
\end{document}